\documentclass[a4paper]{article}
\usepackage[T1]{fontenc}
\usepackage{authblk}
\usepackage{graphicx}

\title{Pattern recognition techniques to reduce backgrounds in the search for the $^{136}$Xe double beta decay with gaseous TPCs}

\author[1]{F.J.~Iguaz}
\author[1]{S.~Cebrian}
\author[1]{T.~Dafni}
\author[1]{H.~Gomez}
\author[1]{D.C.~Herrera}
\author[1]{I.G.~Irastorza}
\author[1]{G.~Luzon}
\author[1]{L.~Segui}
\author[1]{A.~Tom\'as}

\affil[1]{Laboratorio de F\'isica Nuclear y Astropart\'iculas,
Universidad de Zaragoza, Spain}

\begin{document}

\maketitle

\begin{abstract}
The observation of the neutrinoless double beta decay may provide essential information on the nature of neutrinos. Among the current experimental approaches, a high pressure gaseous TPC is an attractive option for the search of double beta decay due to its good energy resolution and the detailed topological information of each event. We present in this talk a detailed study of the ionization topology of the $^{136}$Xe double beta decay events in a High Pressure Xenon TPC, as well as that of the typical competing backgrounds. We define some observables based on graph theory concepts to develop automated discrimination algorithms. Our criteria are able to reduce the background level by about three orders of magnitude in the region of interest of the $^{136}$Xe Q$_{\beta\beta}$ for a signal acceptance of 40\%. This result provides a quantitative assessment of the benefit of topological information offered by gaseous TPCs for double beta decay search, and proves that it is a promising feature in view of future experiments in the field. Possible ideas for further improvement in the discrimination algorithms and the dependency of these results with the gas diffusion and readout granularity will be also discussed.
\\
{\bf Email}: iguaz@unizar.es
\end{abstract}

\section{Motivation}
The reduction of background levels is a key point of double beta decay experiments \cite{Avignone:2008cs}, not only by the use of radiopure materials but also by pattern recognition techniques. This latter topic has been widely treated for Germanium detectors \cite{Gomez:2007hg}, using the pulse shape to discriminate single from multisite events. In Xenon calorimeter approaches, the study made by Gothard in \cite{Wong:1993htw} for a small TPC introduced several ideas, like the presence of a big charge deposit at both ends of the track, and proved the effectiveness of using this distinctive double beta topological signature to discriminate signals from background. The development of these ideas is the aim of \cite{Cebrian:2013sc}, including the full simulation of the signal and background events in a High Pressure Xenon TPC (HPXe TPC) equipped with a pixelized detector. This proceeding is a short summary of this paper and it also includes the latest work on discrimination algorithms.
 
\medskip
In the first section, we present the complete simulation chain of a HPXe TPC with a pixelized readout. The topology of the neutrinoless double beta decay ($\beta\beta0\nu$) of $^{136}$Xe and the expected backgrounds is then compared, based on the physical processes they suffer to deposit energy in the Range of Interest (RoI). From this comparison, two pattern recognition algorithms have been created and are described in the following section. Their rejection power is then presented and we finish with some conclusions and an outlook of possible improvements.

\section{Simulation of a high pressure Xenon TPC}
\begin{figure}[htb!]
\centering
\includegraphics[width=50mm]{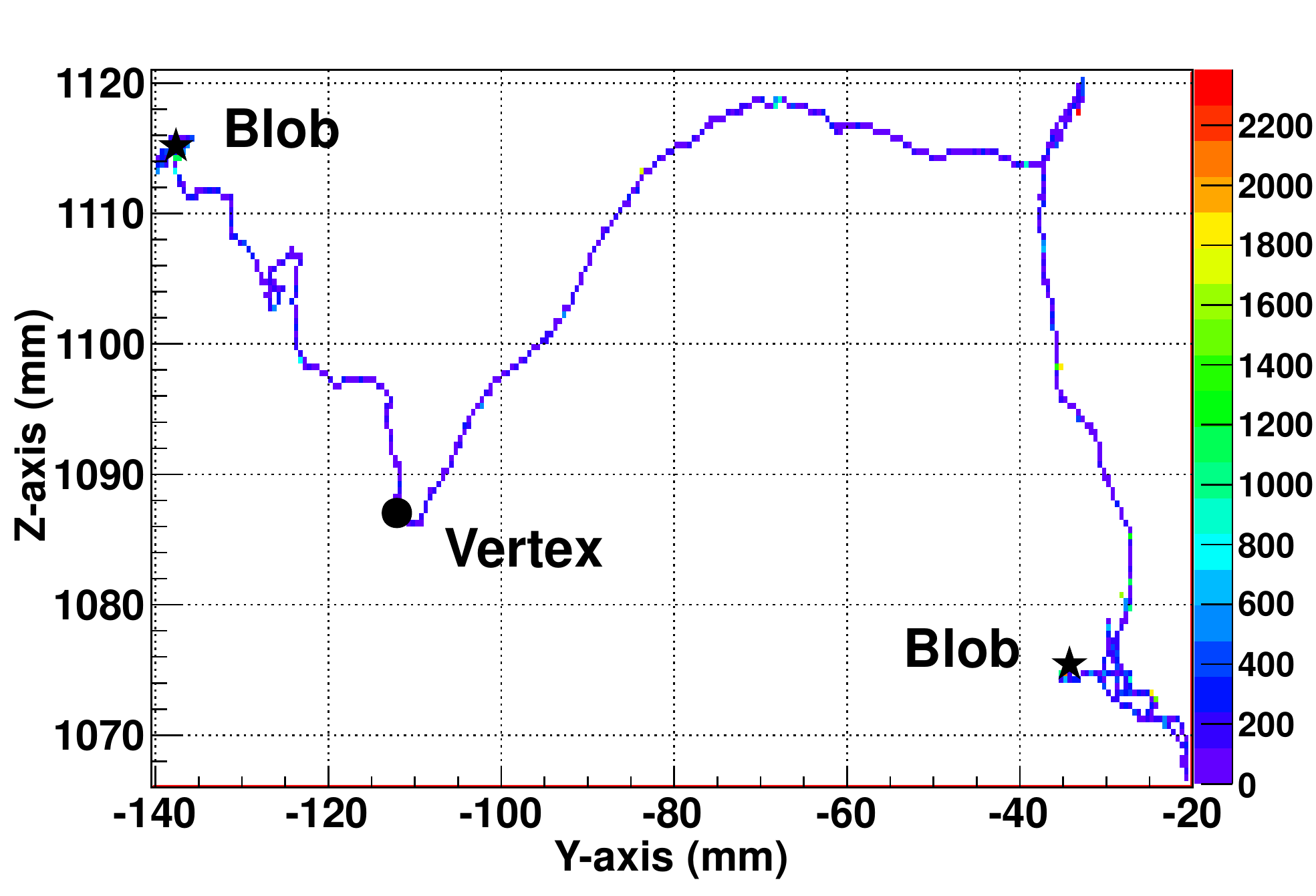}
\includegraphics[width=50mm]{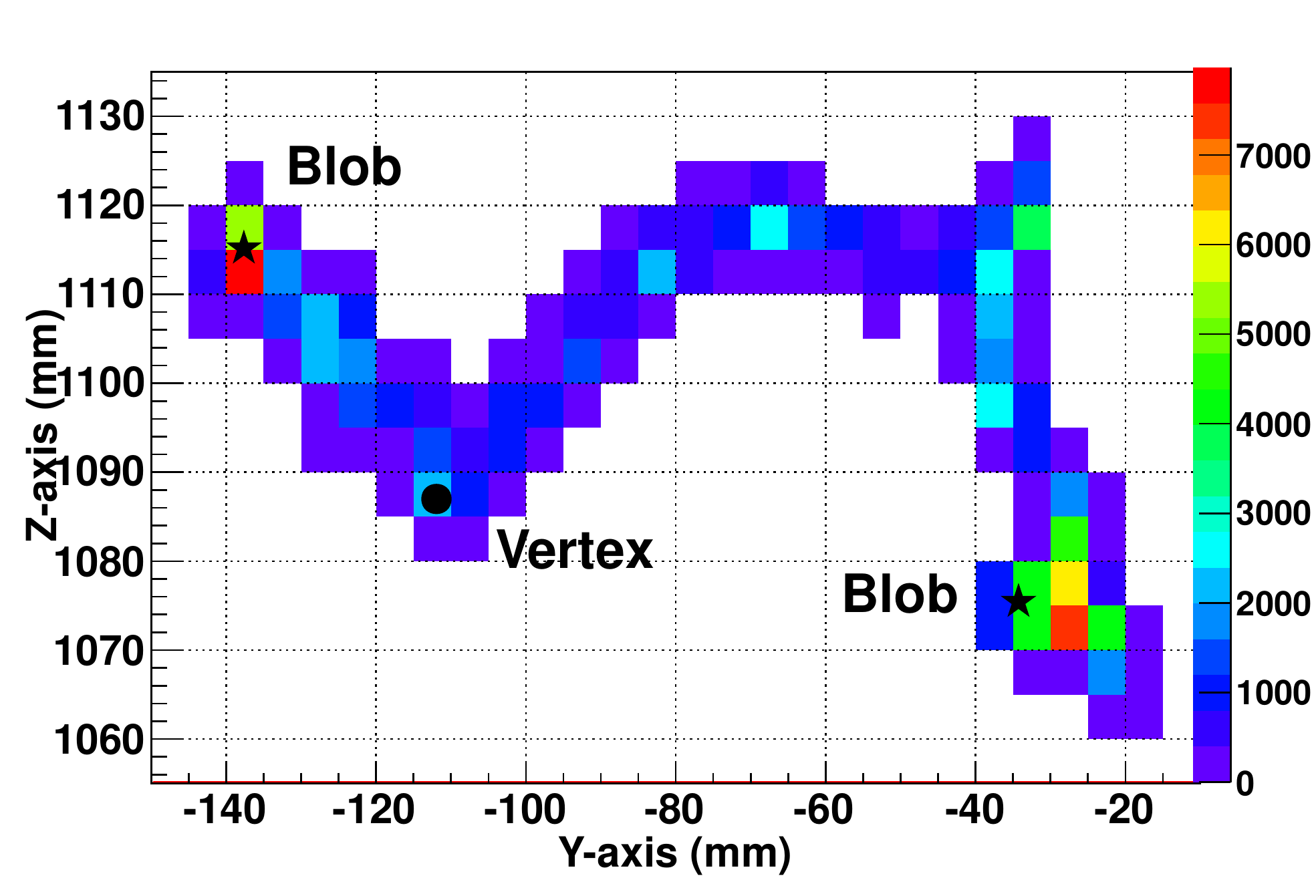}
\includegraphics[width=50mm]{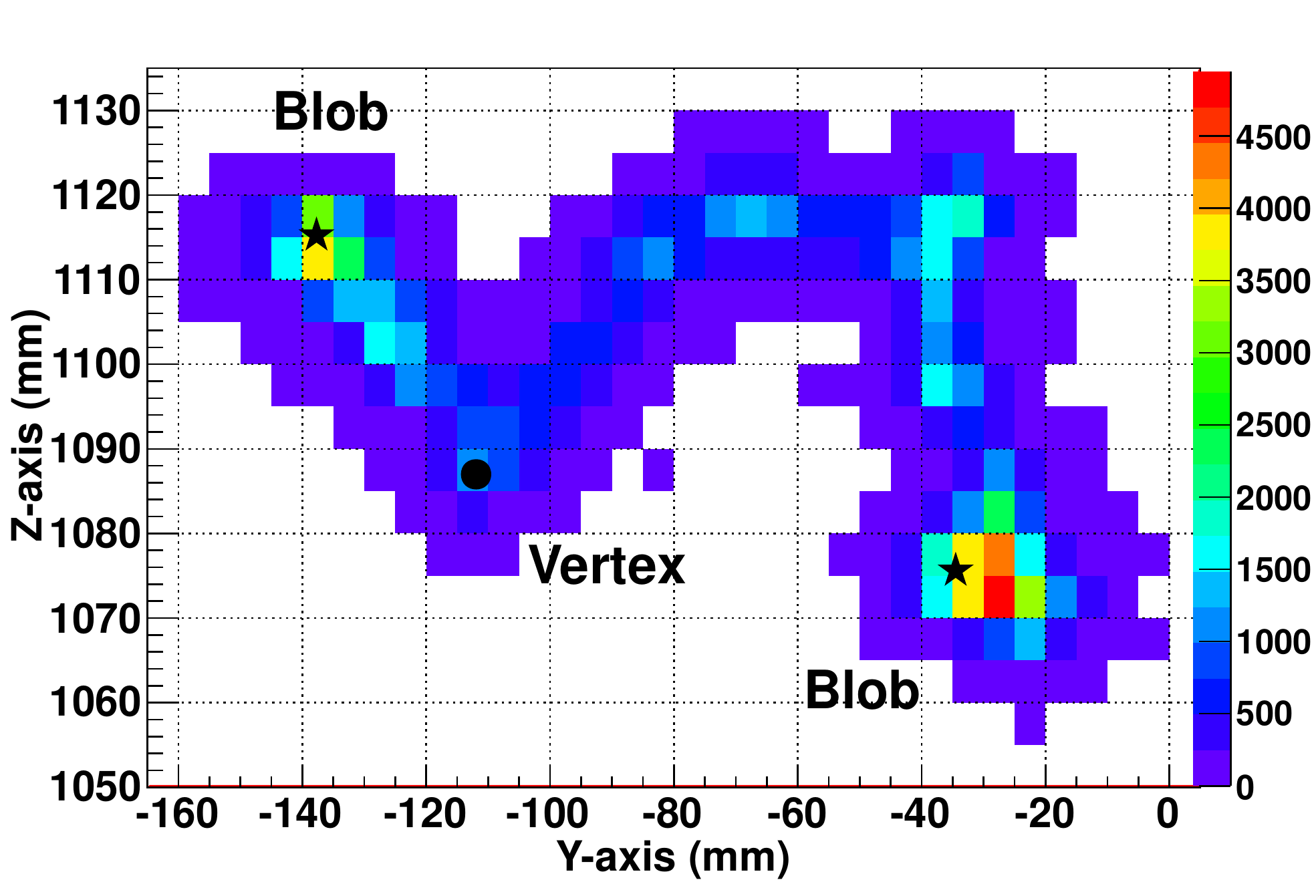}
\caption{\it A simulated $\beta\beta0\nu$ of $^{136}$Xe in a HPXeTPC at 10 bar, directly plotted from Geant4 (left) and as acquired by a pixelized detector in LDXe (center) and HDXe (right). The event consists in two electrons emitted from the black point named vertex. Both electrons show a big energy deposit at the end of their tracks, marked by black stars. A pixel length of 5 mm has exceptionally been set to show the importance of diffusion.}
\label{fig:Simulation}
\end{figure}

The simulation presented here replicates the physical processes involved in the generation of signals in a HPXe TPC. It can be divided into three logical blocks. The first one consists of a Monte Carlo simulation, including all the physical processes involved in the passage of gamma-rays or charged particles through matter. It provides the interaction points and the corresponding energy losses within the gas (figure \ref{fig:Simulation}, left). This part of the simulation is based on Decay0 \cite{Pokratenko:2000op} + GEANT4 simulation packages \cite{Agostinelli:2003sa} and includes a conceptual design of a Xenon TPC. The second block describes the generation of electrons in the gas, the diffusion effects during the drift to the readout plane and the reconstruction of the 3D event by the readout. We have considered a pixel size of $1\times 1$ cm$^2$, a gas pressure of 10 bar and two possible gas diffusion coefficients: low (LDXe, figure \ref{fig:Simulation}, center) and high (HDXe, figure \ref{fig:Simulation}, right). The final block includes the discrimination algorithms and is described in detail below.

\section{Expected backgrounds in the RoI}
In a detailed simulation, several contaminations must be considered to generate the expected background spectrum. However, a first study can reduce the number of contributions to be simulated by the RoI. In our case, we have fixed a RoI between 2400 and 2500 keV, wider than the energy resolution window of current Xenon experiments \cite{Ackerman:2011na}.

\begin{enumerate}
 \item {\bf Two-neutrinos double beta decay mode}: for the half-life of \cite{Ackerman:2011na} and our RoI, the background level is negligible.
 \item {\bf Neutron activation}: the only worrisome isotope is the $^{137}$Xe. It produces $\beta$ emissions with maximum energy of 4173 keV (67\%) and maximum energy of 3717.5 keV plus a $\gamma$ of 455 keV (30\%). This contribution is irrelevant as the computed production (0.6 nuclei/year/kg) is equal to the short half-life of the element (3.8 min).
 \item {\bf Outer contaminations}: the contribution by muons (direct and high energy photons) is negligible in underground laboratories and the background level of gammas is $< 10^{-4}$ keV$^{-1}$ kg$^{-1}$ yr$^{-1}$ with a 25 cm-thick lead shielding.
 \item {\bf Inner contaminations}: both alpha and beta emissions are normally absorbed at the limits of the fiducial volume and can be easily rejected. From gamma emitters, there are only two important contributions: $^{208}$Tl ($\gamma$-line at 2614.5 keV) and $^{214}$Bi ($\gamma$-line at 2447 keV and $\beta$ emissions with maximum energies of 3272 and 824 keV). Note that $^{60}$Co also contributes to the RoI ($\gamma$-lines at 1173 keV and 1332 keV), but these events will be easily identified by their two separated energy deposits.
\end{enumerate}

\section{The topology of signal and background events}
\begin{figure}[htb!]
\centering
\includegraphics[width=50mm]{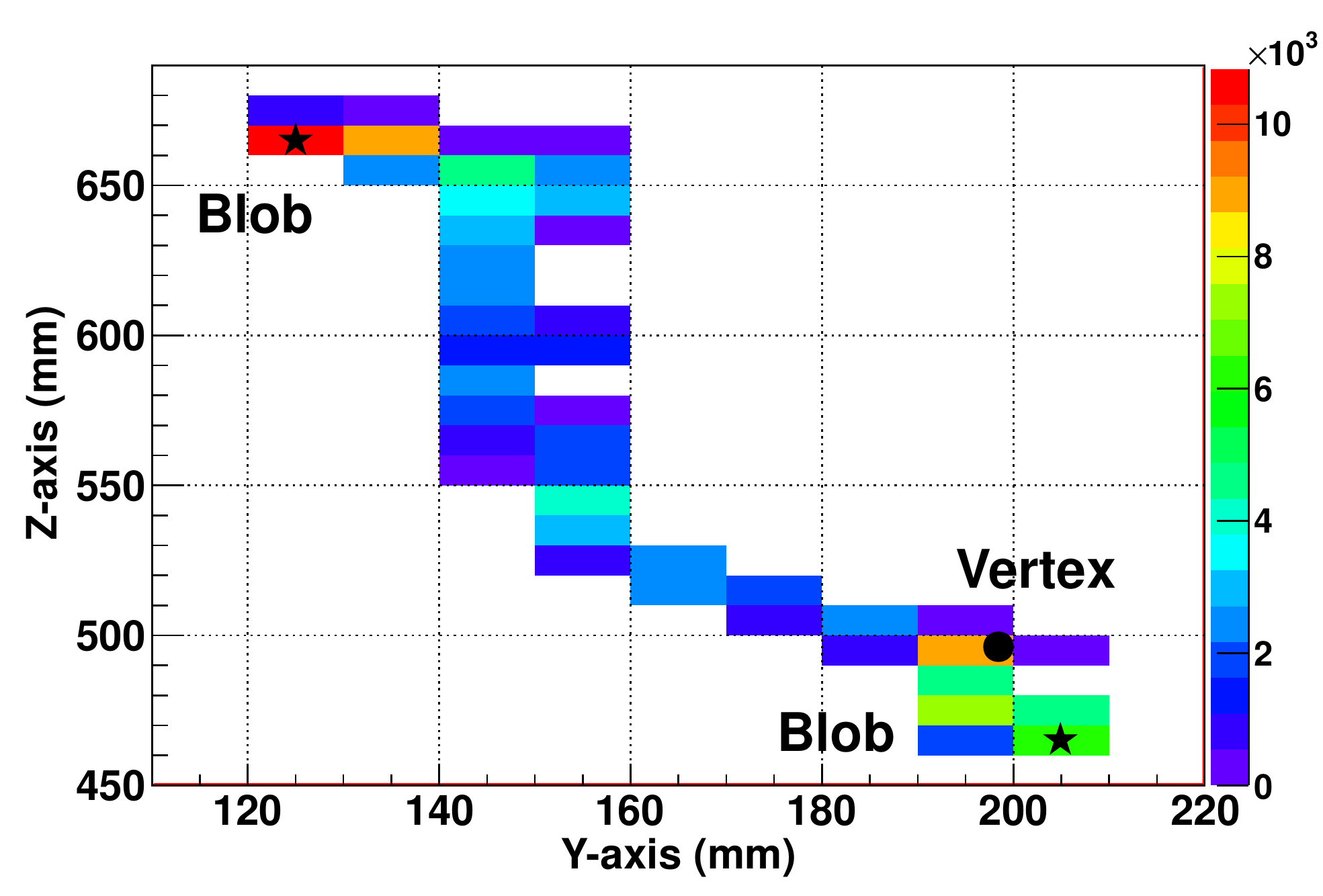}
\includegraphics[width=50mm]{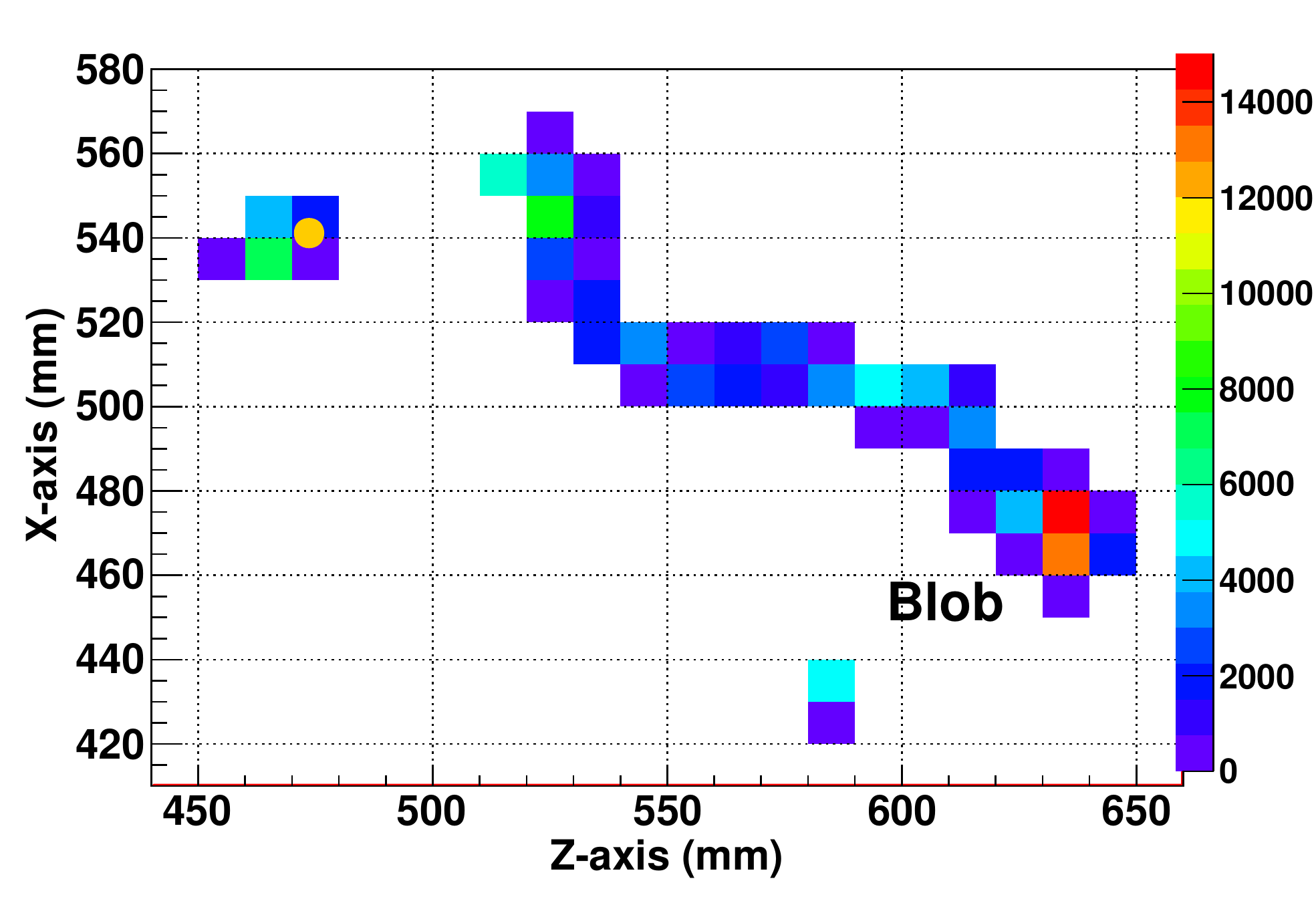}
\includegraphics[width=50mm]{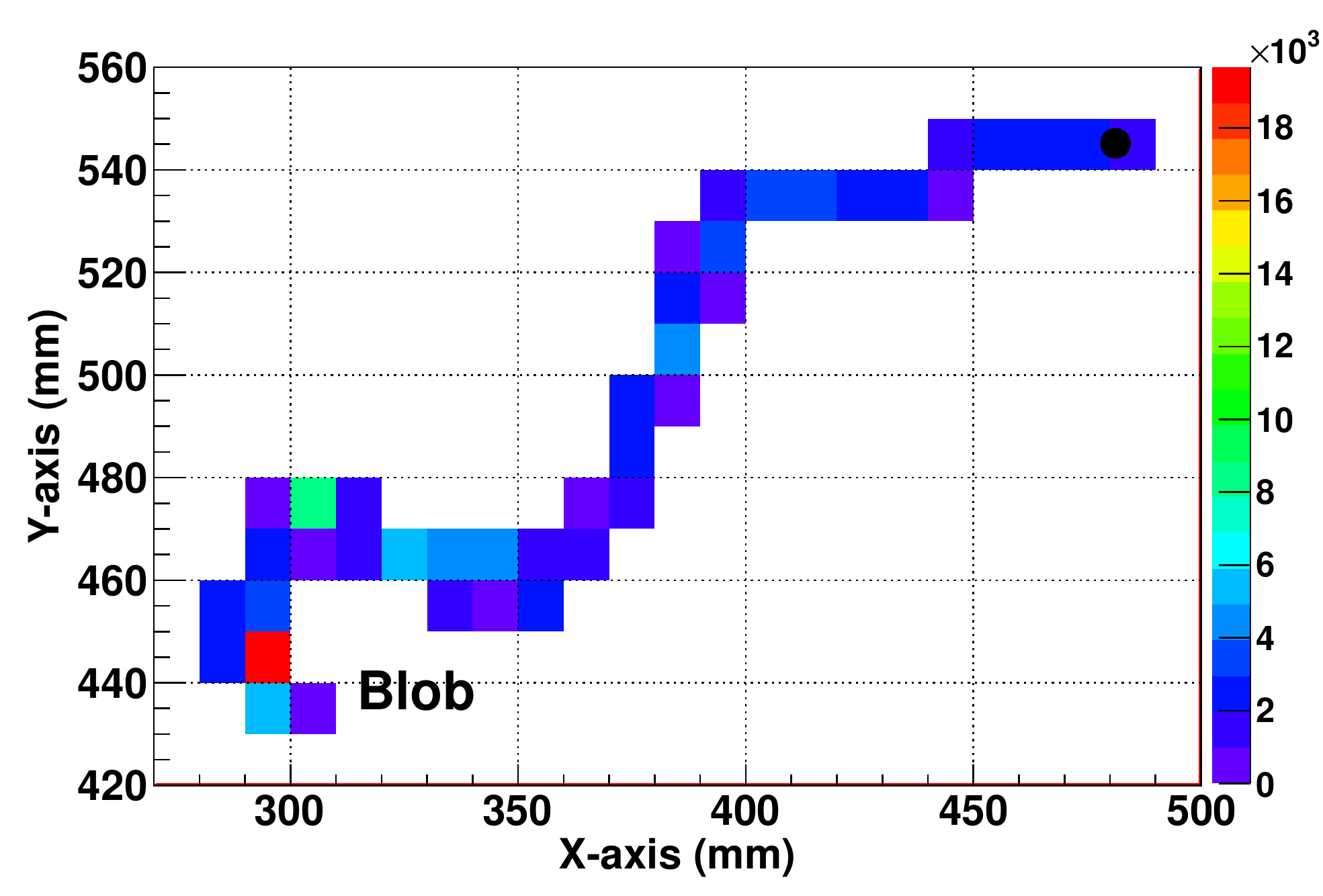}
\caption{\it Left: A simulated $\beta\beta0\nu$ of $^{136}$Xe detected in a HPXe TPC. The event consists in two electrons of 1.11 and 1.35 MeV emitted from the black point named vertex. Both electrons show a big energy deposit (or {\it blo}b) at the track's end, marked by a black star. Center: An event in the RoI generated by $^{208}$Tl. There are four tracks and the main one shows a big charge deposit at only one end. Right: A single-track event in the RoI generated by $^{214}$Bi. Only one of the track's end show a blob.}
\label{fig:Discrimination}
\end{figure}

The topological features of background events in the RoI are different depending on the physical processes that suffer. They are also different from the expected signal, which is a continuous staggered track with a high energy deposit at both ends (figure \ref{fig:Discrimination}, left). We will show in this section these differences, studying both $^{208}$Tl and $^{214}$Bi.

\medskip
Simple physical processes (photoelectric, Compton scattering or pair production) of the $^{208}$Tl $\gamma$-line cannot generate an event in the RoI, but by three more complex processes. The first one is a Compton scattering in an external volume followed by a photoelectric absorption or Compton scattering in the gas. The second one is a multi-Compton scattering in the gas with a partial energy loss in form of low energy gammas. In these two cases, there are two energy deposits spatially separated (figure \ref{fig:Discrimination}, center), whereas the expected signal will generally have only one track. This fact does not appear in the third process, which consists in a photoelectric absortion followed by the emission of bremstrahlung photons which escape from the TPC. However, it will normally show only a big energy deposit at one track's end.

\medskip
Most background events of $^{214}$Bi consist in the photoelectric absorption of the intense $\gamma$-line at 2448 keV. These events will have a single component but also only a big energy deposit at one track's end (figure \ref{fig:Discrimination}, right). Other contributions are the Compton scattering of less intense higher-energy $\gamma$-lines and the $\beta$ emissions from the vessel's inner walls. In those cases, there would be either multiple tracks or an energy deposit at the fiducial volume limits.

\medskip
In summary, two topological features have been identified, which are not present in background events: the absence of energy deposits far from the main component (or {\it trac}k) and the presence of a big charge deposit (generally called {\it blob}) at both track ends. The discrimination algorithms based on these features are described in the following section.

\section{Discrimination algorithms based on Graph Theory}
\subsection{Track/Component search}
Each event is firstly transformed to a mathematical graph: pixels are identified with vertex and adjacent pixels are linked by a segment. We then use a classical method of Graph Theory to find tracks or components \cite{Bollobas:1998bb}. Most of background events are multiple-track but also only $\approx$60\% of signal events are single-track, due to the emission of xenon x-rays (energy $\approx$ 30 keV) or low energy bremsstrahlung photons. These two effects produce a dramatic reduction of signal efficiency if only single-track events are selected. For this reason and based one the factor of merit, events with one extra short track of less than 100 keV are also accepted.

\subsection{Two-blobs identification}
This algorithm is based on the method to find the longest track-line in a graph \cite{Bollobas:1998bb}. As a first step, several blob candidates are identified setting an energy threshold of 150 keV. Then, two track-lines are generated: one between two blob candidates and another between one blob and a normal pixel. The algorithm choses the first line except if the second one is 30\% longer than the first track-line. This criterium is a compromise to recognize:
\begin{enumerate}
 \item Signals with two blobs or those where one of the electrons has not enough energy to generate a blob.
 \item Backgrounds generally show one blob but random energy deposits (like $\delta$-rays, bremsstrahlung photons or low Compton interacting close to the main track) may create additional blobs at the main track.
\end{enumerate}
Finally, the charge of both blobs is calculated considering a fixed radius around them. The blob with less charge is a good parameter to select signal events, as already shown by Gothard collaboration \cite{Wong:1993htw}. In our case, we have set a threshold of 440 keV. However, background events with fake blobs at the main track could not be rejected as the track-line will generally end at those charge deposits. In those cases, an important part of the charge will be out of this track-line and the track-to-total-charge ratio (named {\it track coverage}) can be used to reject them.

\subsection{Background rejection power and the lattest developments}
\begin{figure}[htb!]
\centering
\includegraphics[width=75mm]{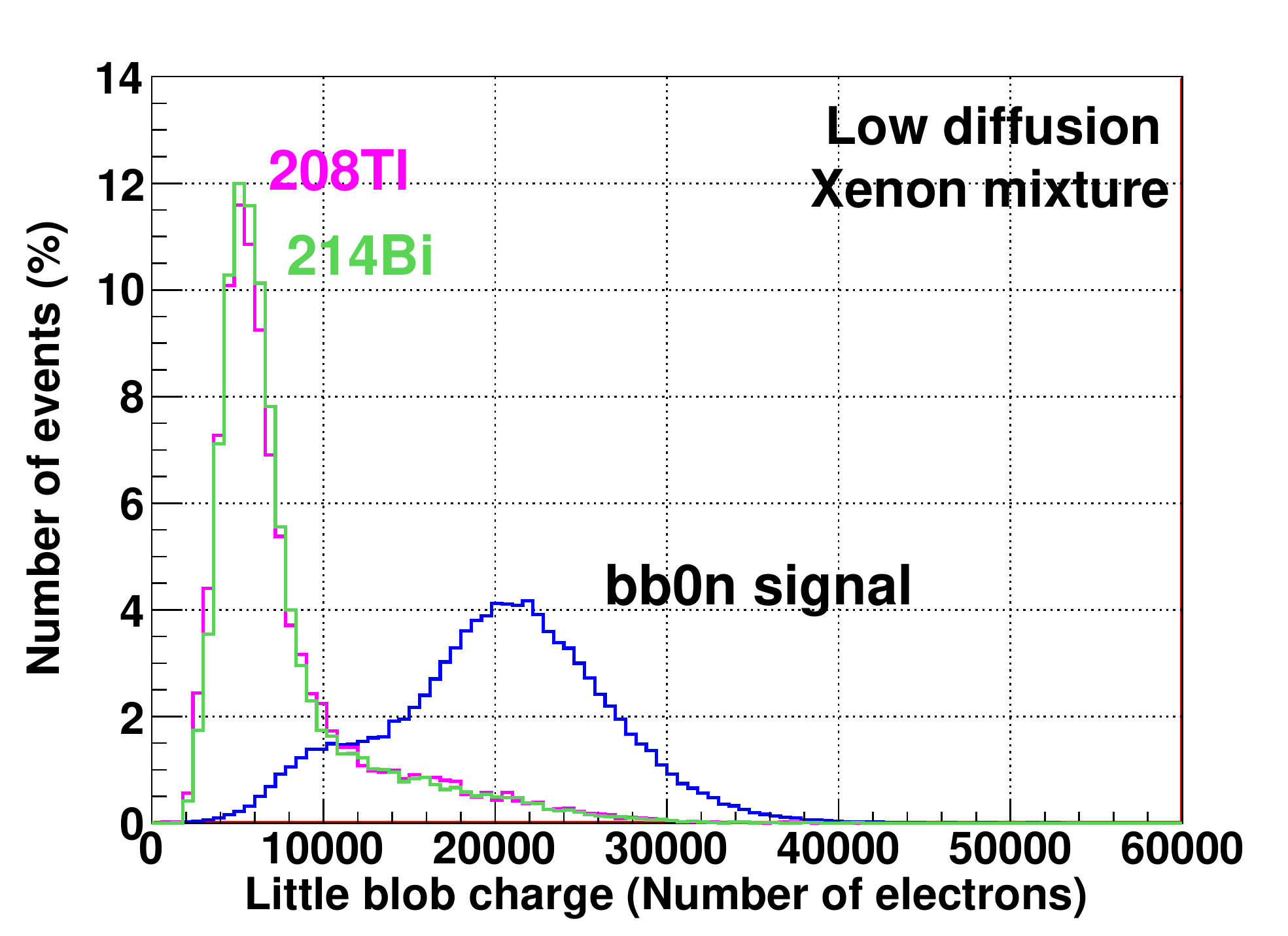}
\includegraphics[width=75mm]{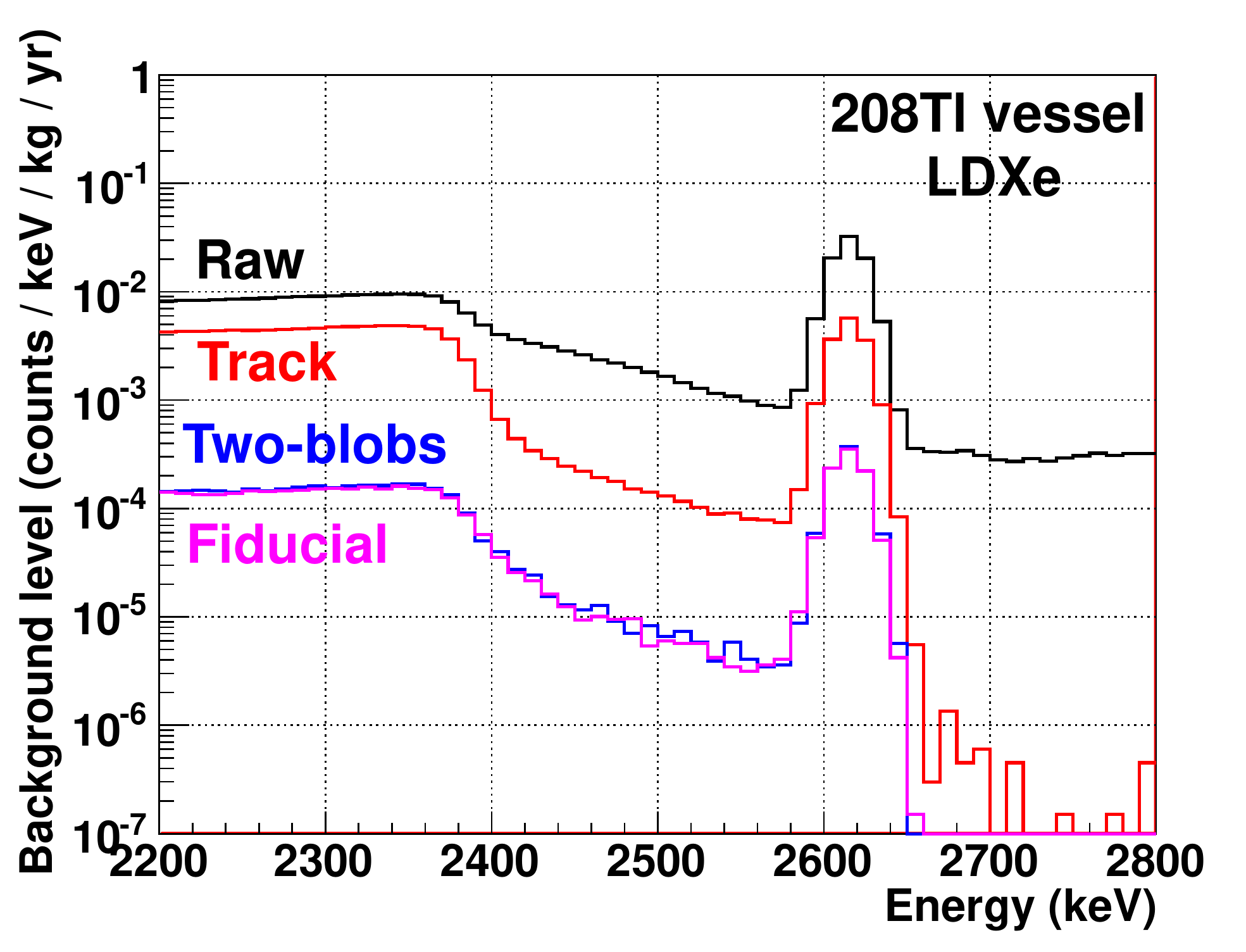}
\caption{\it Left: Charge distribution of the little blob in LDXe generated by $\beta\beta$0$\nu$ (blue line) and $^{208}$Tl (magenta line) and $^{214}$Bi (green line) vessel contaminations. Events are in the RoI and have been previously selected by the track criterium. Right: Energy spectrum generated by $^{208}$Tl events emitted from the vessel (with an activity of 10 $\mu$Bq/kg) in a LDXe after the succesive application of the selection criteria: initial (black line), track (red line), two-blobs (blue line) and fiducial volume (magenta line).}
\label{fig:LittleCharge}
\end{figure}

Fixing a 40\% signal efficiency, track criterium rejects mainly $^{208}$Tl events (4.8-7.1\% background efficiency\footnote{Defined as the percentage of background events that survive a selection criterium.}, depending on the vertex), as events in the RoI suffer several physical processes. The efficiency for $^{214}$Bi is modest (20\%) as most events are generated by a photoelectric absorption. The topological criterium shows an efficiency of 12\% for LDXe, near the value of 8.6\% published by Gothard \cite{Wong:1993htw}, while the HDXe case is a factor 3 worse due to a modest blob identification. Current work is focused on a better determination of blob positions and the elimination of fine-tuning variables from our algorithms. Preliminary results show a cleaner separation of the charge distribution of the little blob (figure \ref{fig:LittleCharge}, left), that respectively improves the background rejection efficency to 6.0 and 9.2\% in LDXe and HDXe, independently of the vertex position and the type of contamination. As a result, the background level is reduced more than two orders of magnitude in the RoI (figure \ref{fig:LittleCharge}, right).

\section{Conclusions and prospects}
A detailed simulation of a HPXe TPC with a pixelized readout for a double beta decay experiment of $^{136}$Xe has been presented. We have compared the topology of signal and backgrounds to identify two key topological features of signals: a single track and one big charge deposit at both ends. Two pattern recognition algorithms based on Graph Theory have been created to identify tracks and to detect blobs, with good results. Current work is focused on a better determination of blob position, the elimination of fine-tuning variables and the study of background rejection dependence on other TPC features like pressure and pixel size.

\section{Acknowledgments}
We are grateful to our colleagues of the groups of the University of Zaragoza, CEA/Saclay as well as the NEXT and RD-51 collaborations. We acknowledge support from the European Commission under the European Research Council T-REX Starting Grant ref. ERC-2009-StG-240054 of the IDEAS program of the 7th EU Framework Program. We also acknowledge support from the Spanish Ministry of Economics and Competitiveness (MINECO), under contracts ref. FPA2008-03456 and FPA2011-24058, as well as under the CUP project ref. CSD2008-00037
and the CPAN project ref. CSD2007- 00042 from the Consolider-Ingenio 2010 program of the MICINN. Part of these grants are funded by the European Regional Development Fund (ERDF/FEDER). F.I. acknowledges support from the Eurotalents program.


\begin{thebibliography}{9}
\bibitem{Avignone:2008cs}
  F.~T.~Avignone, S.~Elliott \& J.~Engel,
  {\it Rev. Modern. Phys.} {\bf 80} (2008) 481.
\bibitem{Gomez:2007hg}
  H.~G\'omez {\it et al.},
  {\it Astropart. Phys.} {\bf 28} (2007), 435.
  M.~Agostini {\it et al.},
  {\it JINST} {\bf 6} (2011) P04005.
\bibitem{Wong:1993htw}
  H.~T.~Wong {\it et al.},
  {\it Nucl.\ Instrum.\ Meth.\  A} {\bf 329} (1993) 163.
\bibitem{Cebrian:2013sc}
  S.~Cebrian {\it et al.},
  article in preparation.
\bibitem{Pokratenko:2000op}
  O.A.~Pokratenko {\it et al.},
  {\it Phys. At. Nucl.} {\bf 63} (2000) 1282.
\bibitem{Agostinelli:2003sa}
  S.~Agostinelli {\it et al.},
  {\it Nucl.\ Instrum.\ Meth.\  A} {\bf 506} (2003) 250.
  J.~Allison {\it et al.},
  {\it IEEE Trans. Nucl. Sci.} {\bf 53} (2006) 270.
\bibitem{Ackerman:2011na}
  N.~Ackerman {\it et al.},
  {\it Phys. Rev. Lett.} {\bf 107} (2011) 212501.
  A.~Gando {\it et al.},
  {\it Phys. Rev. C} {\bf 85} (2012) 045504.
\bibitem{Bollobas:1998bb}
  B.~Bollob\'as, 
  Modern Graph Theory,
  Springer-Verlag, 1998.
\end{thebibliography}
\end{document}